\documentclass[reprint,superscriptaddress,aps,nofootinbib,prd]{revtex4-2}

\usepackage{amsmath}
\usepackage{graphicx}
\usepackage[colorlinks=true,linkcolor=blue,citecolor=black,urlcolor=black]{hyperref}
\usepackage{orcidlink}
\usepackage{cleveref}

\crefname{equation}{Eq.}{Eqs.}
\Crefname{equation}{Eq.}{Eqs.}
\crefname{figure}{Fig.}{Figs.}
\Crefname{figure}{Fig.}{Figs.}

\hypersetup{
  colorlinks=true,
  urlcolor=[rgb]{0.2,0.2,0.7},
  citecolor=[rgb]{0.2,0.2,0.7},
  linkcolor=[rgb]{0.2,0.2,0.7}
}

\begin{document}

\title{Non-singular cosmology from effective LQG-inspired black holes}

\author{V. F. C. Belo\orcidlink{0009-0003-3156-9873}}
\thanks{\href{mailto:vitoriacampos@ufba.br}{vitoriacampos@ufba.br}}
\author{F. G. S. Menezes\orcidlink{0009-0002-1170-8090}}
\thanks{\href{mailto:humberto@ufba.br}{menezes.filipe@ufba.br}}
\author{S. M. Rupf\orcidlink{0009-0005-2596-2980}}
\thanks{\href{mailto:sthefannymoreira@ufba.br}{sthefannymoreira@ufba.br}}

\affiliation{Group of Gravitation and Cosmology, Instituto de F\'{\i}sica, Universidade Federal da Bahia, 40210-340, Salvador, BA, Brazil}
\author{I. P. R. Baranov\orcidlink{0000-0003-4681-3042}}
\thanks{\href{mailto:iuribaranov@gmail.com}{iuribaranov@gmail.com}}
\affiliation{Group of Gravitation and Cosmology, Instituto de F\'{\i}sica, Universidade Federal da Bahia, 40210-340, Salvador, BA, Brazil}
\affiliation{Instituto Federal de Educa\c c\~ao, Ci\^encia e Tecnologia da Bahia, 40301-015, Salvador, BA, Brazil}

\author{H. A. Borges\orcidlink{0000-0001-7553-8540},}
\thanks{Corresponding author: \href{mailto:humberto@ufba.br}{humberto@ufba.br}}
\affiliation{Group of Gravitation and Cosmology, Instituto de F\'{\i}sica, Universidade Federal da Bahia, 40210-340, Salvador, BA, Brazil}

\date{\today}

\begin{abstract}

We propose a non-singular primordial cosmology derived from matching an effective Loop Quantum Gravity-inspired black hole geometry to an expanding Friedmann--Lema\^{i}tre--Robertson--Walker universe via Darmois--Israel junction conditions. The resulting modified Friedmann equation replaces the singularity with a quantum bounce in the Planckian regime. Mapping the surface energy-momentum tensor of the dynamic thin shell onto a homogeneous scalar field reveals a transient phantom-like behavior. For any equation of state of the shell satisfying $\omega > -1$, its effective energy density is diluted with cosmic expansion. Extending the framework to a two-scalar-field system provides a mechanism to bridge the energy gap between the Planckian bounce and the sub-Planckian scale. The pre-inflationary phase is brief, enabling the second scalar field to take over to drive the standard slow-roll inflation.

\end{abstract}

\maketitle

\section{Introduction} 

The inflationary scenario is a description of the primordial universe that offers a solution  to well-known cosmological problems \cite{Guth, Sta, AA}, as well as providing a mechanism for the generation of quantum fluctuations in the inflaton field that act as seeds for the anisotropies observed in the Cosmic Microwave Background (CMB) radiation \cite{P,Aghanim} and for the large-scale structure of  the universe \cite{M}. However, standard inflation is described within semiclassical General Relativity (GR) in such a way that the initial singularity remains \cite{AB}, which is expected to be resolved by a fundamental theory of quantum gravity \cite{Ashtekar:2003hd,Ashtekar:2009ov,LQC:2026}.

Among the candidates for quantum gravity, Loop Quantum Gravity (LQG) \cite{lewandowski, livros1, livros3} provides a non-perturbative framework where quantum corrections regularize singularities. In the context of black hole spacetimes, polymerization procedures \cite{PRL, AOS, Gambini, espanhois, bascos2} replace the central singularity with an internal transition surface at a minimum radius $x_0$ \cite{Fernando, Filipe}, interpolating between black and white hole interiors. From an effective phenomenological viewpoint, this LQG-inspired geometry can be interpreted as standard general relativity coupled to an anisotropic effective fluid \cite{DAC}.  At low curvatures, the energy density and pressures of this effective fluid are negligible. However, as the curvature approaches Planckian scales, the effective density and pressures violate the classical energy conditions of GR and generates the repulsive effects necessary to resolve the singularity \cite{worm}.

Beyond the singularity issue, an important question concerns initial conditions and the energy hierarchy between primordial scales. Observational constraints on the CMB data favor inflationary potentials with a nearly flat plateau \cite{Aghanim, SA}, typically on the Grand Unification (GUT) scale, with an energy density limited by $\sim (10^{16}\text{ GeV})^4$. If the universe emerges directly from a Planck scale $\sim (10^{19}\text{ GeV})^4$, the plateau density is approximately $10^{12}$ times lower than the Planckian density, and an intermediate dynamical stage is required to dilute the energy density before the onset of conventional slow-roll inflation, which requires a fine-tuning of initial homogeneity across independent causal patches \cite{Maeda1989, Lazarides1998, Boutaleb2001, YamaguchiYokoyama2001, Yamaguchi2001, Panagiotakopoulos, Dimopoulos, Linde}. Therefore, a primordial mechanism is needed to bridge this energy gap.

In this work, we construct a cosmological model that resolves the initial singularity and establishes pre-inflationary conditions at the Planck scale. The geometric construction relies on matching an effective LQG-inspired black-hole geometry to an expanding Friedmann--Lema\^{i}tre--Robertson--Walker (FLRW) spacetime through Darmois--Israel (DI) junction conditions \cite{Dar, Is, Lan}. The discontinuity of the extrinsic curvature across the matching hypersurface gives rise to a dynamical thin shell with surface energy density $\sigma$ and surface pressure $p$. As recently shown in the context of primordial black holes acting as cosmic expansion accelerators \cite{K}, matching a repulsive interior geometry to a cosmological background can drive an inflationary phase. In fact, for a spatially flat FLRW this framework leads to a modified Friedmann equation in which the primordial classical singularity is replaced by a quantum bounce at the minimum radius $x_0$.

By modeling a homogeneous distribution of black holes within a ``Swiss Cheese" model \cite{SW, G, F}, the thin shells can be mapped onto an effective fluid and consequently onto a homogeneous scalar field. In the vicinity of the minimum radius, the scalar field exhibits a negative kinetic term such that the field is interpreted as a transient phantom-like field. We show that its effective energy density dilutes rapidly with expansion. Thus, to connect the Planck scale to the inflationary scale, we extend the model to a two-scalar-field system. In this setup, the transient phantom-like field drives the Planckian dynamics and rapidly dilutes, enabling the canonical field to take over at lower energies. 

The paper is organized as follows. In \cref{secII}, we review the effective LQG-inspired black hole geometry. In \cref{secIII}, we present the Darmois--Israel junction conditions used to match this local geometry to an expanding FLRW universe, deriving the modified Friedmann equation and the conditions required for a bounce to occur. In \cref{secIV}, we map the surface dynamics of the thin shell onto a scalar field, analyzing the transient phantom-like behavior near the bounce, the asymptotic decay, and the two-scalar field mechanism that bridges the Planckian regime to slow-roll inflation. Finally, in \cref{secV}, we summarize our main conclusions. The derivation of the continuity equation for the shell and the junction conditions derived via the Hamiltonian formalism are detailed in Appendices \ref{apd:A} and \ref{apd:B}, respectively.
 
\section{Effective LQG-inspired black hole metric \label{secII}}

We begin by considering the spacetime metric with spherical symmetry as
\begin{equation}
ds^2 = - N^2 dT^2 + \frac{p_b^2}{p_c} d\tilde x^2 + p_c d\Omega^2,
\end{equation}
where the lapse function is $N = \gamma \sqrt{p_c}/b$. Within this framework, the triads $(p_b, p_c)$ and their corresponding connections $(b, c)$ satisfy the Poisson structure relations $\{b, p_b\} = \gamma$ and $\{c, p_c\} = 2\gamma$, where $\gamma$ is the Barbero--Immirzi parameter. Note that the triad component $p_c$ is related to the physical radius of the two-sphere metric. The classical Hamiltonian that governs the dynamics is expressed as
\begin{equation}
    H_{\rm cl}=-\frac{1}{2 G \gamma}\left[\left(b+\frac{\gamma^2}{b} \right) p_b + 2c p_c    \right].
\end{equation}

The effective model can be obtained by employing the polymerization procedure \cite{Gambini}
\begin{equation}\label{eq:polymerisation}
b \rightarrow \frac{\sin\left(\delta_b b\right)}{\delta_b},\quad p_b \rightarrow \frac{p_b}{\cos(\delta_b b)},
\end{equation}
with regularization factor
\begin{equation}\label{eq:regularisation}
\frac{\cos(\delta_b b)}{\sqrt{1+\gamma^2\delta_b^2}},
\end{equation}
in the classical Hamiltonian, where $\delta_b$ is a parameter that controls quantum corrections. 
Then, the solutions of the equations of motion give rise to the static metric \cite{ espanhois, bascos2, Fernando} 
\begin{equation} \label{metricS}
   ds^2 = - \bigg(1-\frac{2m}{x}\bigg) d\tau^2 + \frac{dx^2}{\left(1-\frac{2m}{x}\right)\left(1-\frac{x_0}{x}\right)} + x^2 d\Omega^2,
\end{equation}
where $m$ is a constant of motion that is identified with the black hole mass, $x_0$ represents the minimum radius related to quantum corrections, and $d\Omega^2\equiv d\theta^2+\sin^2\theta d\phi^2$. We adopt $c=G=\hbar=1$ units throughout this work. In this effective geometry, the classical singularity is replaced by a transition surface 
interpolating between black and white hole interiors of equal masses, with horizons located at 
$2m$. We always have $2m>x_0$, that is, the transition
surface lies inside the horizon \cite{Fernando}. The black hole evaporates down until it reaches the minimum mass
$x_0/2$ \cite{Filipe, Belfaqih2025, AlonsoBardaji2025}, in contrast to the classical picture, in which complete evaporation is expected. The area of the transition surface is equal to the LQG area gap, so that $x_0 =\sqrt{\sqrt{3}\gamma}$, where $\gamma$ is the Barbero--Immirzi parameter. Throughout our numerical evaluations, we adopt the standard value $\gamma =\sqrt{3}/6$ \cite{Pigozzo_2021}. In the regime $x\gg x_0$, the effective metric approaches the Schwarzschild geometry.

\section{Junction conditions and modified cosmological dynamics \label{secIII}}

To construct a cosmological model, we match two metrics along a common timelike hypersurface $\Sigma$ that separates our four-dimensional spacetime into an interior region ($\mathcal{M}^-$), described by the black hole metric in the form
\begin{equation}\label{metric1}
ds^2_{-}=-f(x)d\tau^2+g(x)dx^2+x^2d\Omega^2,
\end{equation}
and an exterior region ($\mathcal{M}^+$), described by the FLRW metric
\begin{equation}\label{FLS}
ds^2_{+}=-dt^2+a^2(t)\bigg[\frac{dr^2}{1-\kappa r^2}+r^2d\Omega^2\bigg],
\end{equation}
where $a(t)$ is the scale factor and $\kappa$ is the spatial curvature. The conditions for matching the two metrics are referred to as the DI junction conditions. The first junction condition tells us that the induced metric on the hypersurface must be continuous, while the second condition implies that the extrinsic curvature may be discontinuous \cite{Dar, Is, Lan}. This discontinuity corresponds to a layer of energy density and pressure. 

For a given spacetime metric $g_{\mu\nu}$ with global coordinates $X^\mu$, its induced metric obtained by projecting each component onto a hypersurface $\Sigma$ with coordinates $y^i$ is given by \footnote{In a spacetime, the hypersurface $\Sigma$ is 3D with intrinsic coordinates $y^i = (t, \theta, \phi)$, where $t$ is the cosmic time, while $\tau$ denotes the black hole coordinate time. Spatial cross-sections of $\Sigma$ at constant $t$ correspond to 2D spherical thin shells.}
\begin{equation}
h_{ij}=g_{\mu\nu}\frac{\partial X^{\mu}(y)}{\partial y^{i}}\frac{\partial X^{\nu}(y)}{\partial y^{j}}\ ,
\end{equation}
such that imposing the first junction condition we obtain 
\begin{flalign}
\label{factor}
&x(t)=a(t)r_{\Sigma},\\
\label{const}
&-f(x)\dot\tau^2+g(x)\dot x^2=-1, 
\end{flalign}
where a dot means derivative with respect to the cosmological time $t$. The first equation states that the radial coordinate of the black hole, $x(t)$, is directly proportional to the scale factor of the universe, $a(t)$. This connects the local dynamics of the hypersurface to the evolution of the universe, where $r_{\Sigma}$ is the comoving radius in the FLRW coordinates. The second equation ensures that the trajectory of $\Sigma$ is timelike. 

To fully describe the geometry of $\Sigma$, we obtain its normal vectors. Since a normal vector is orthogonal to all tangent vectors at any point on the hypersurface, and the angular coordinates ($\theta, \phi$) are common to both metrics, the orthogonality condition is given by
\begin{equation}
\label{orthcond}
\eta_{\tau}\frac{d\tau}{d t}+\eta_{x}\frac{d x}{dt}=0,
\end{equation}
where $\eta_{\tau}$ and $\eta_{x}$ are the components of the unit normal vector on $\Sigma$ in the black hole's coordinate system. Furthermore, since the hypersurface is timelike, the unit normal vector is normalized by the condition
\begin{equation}\label{unit}
\frac{\eta_x^2}{g(x)}-\frac{\eta_{\tau}^2}{f(x)}=1.
\end{equation}
Combining \cref{factor,const,orthcond,unit}, it is found that the components of the normal vector are 
\begin{equation}
\eta_{\mu}=\pm\sqrt{fg}\bigg(\dot x, -\dot\tau, 0, 0\bigg),
\end{equation}
where the sign of $n_\mu$ depends on the direction of the normal vector. In the DI formalism, the unit normal vector is defined by convention to point from the interior region $\mathcal{M}^-$ to the exterior region $\mathcal{M}^+$.  Therefore, the choice of the negative sign identifies the LQG-inspired black hole as the interior region and the expanding FLRW geometry as the exterior region. In FLRW coordinates, the corresponding unit normal vector is straightforward to obtain, having only a non-zero radial component
\begin{equation}
\tilde\eta_{\mu}=\bigg(0, \frac{a}{\sqrt{1-\kappa r^2}}, 0, 0\bigg).
\end{equation}

The surface stress-energy tensor can be calculated using the Lanczos equation, which is given by
\begin{equation}\label{lanczos}
S_{ij}=-\frac{1}{8\pi}([K_{ij}]-h_{ij}[K]),
\end{equation}
where $K_{ij}$ is the extrinsic curvature given by
\begin{equation}
K_{ij}=-n_{\gamma}\bigg(\frac{\partial^2X^{\gamma}}{\partial y^i\partial y^j}+\Gamma^{\gamma}_{\alpha\beta}\frac{\partial X^{\alpha}}{\partial y^i}\frac{\partial X^{\beta}}{\partial y^j}\bigg),
\end{equation}
and we define square brackets $[K_{ij}]\equiv K_{ij}^{+}-K_{ij}^{-}$ to denote the jump from the black hole ($\mathcal{M}^-$) to the FLRW ($\mathcal{M}^+$) side across the hypersurface. Therefore, if the extrinsic curvature is discontinuous, the hypersurface can typically have its own surface energy-momentum tensor, $S_{ij}$.

In the black-hole region, the only non-zero extrinsic curvature components at $x(t)=a(t)r_{\Sigma}$ are
\begin{flalign}
\label{kt}
&K_{tt}^{-}=\sqrt{fg}\bigg[\dot x\ddot\tau-\dot\tau\ddot x-\frac{f'}{2g}\dot\tau^3+\bigg(\frac{f'}{f}-\frac{g'}{2g}\bigg)\dot\tau\dot x^2\bigg],\\
\label{phi1}
&K_{\phi\phi}^{-}=K_{\theta\theta}^{-}\sin^2\theta= x\dot\tau\sqrt{\frac{f}{g}}\sin^2\theta,
\end{flalign}
where a prime denotes the derivative with respect to $x$, while in the FLRW, 
\begin{flalign}
\label{k+tt}
&K_{tt}^{+}=0,\\
\label{k+phiphi}
&K_{\phi\phi}^{+}=K_{\theta\theta}^{+}\sin^2\theta=a(t)r_{\Sigma}\sqrt{1-\kappa r_{\Sigma}^2}\sin^2\theta.
\end{flalign}
To properly treat the matching hypersurface as a physical thin shell endowed with surface energy density and pressure, we model its surface stress-energy tensor, in analogy with a perfect fluid, as
\begin{equation}\label{perf}
S_{ij}=(\sigma +p)u_iu_j+ph_{ij},
\end{equation}
where $h_{ij}=\textrm{diag}(-1, x^2, x^2\sin^2\theta)$ in the comoving frame.  Using \cref{kt,phi1,k+tt,k+phiphi} and conditions ($\ref{factor}$) and ($\ref{const}$) in  \cref{lanczos} and comparing to \cref{perf}, we obtain the following equations, respectively, for the evolution of surface energy density and surface pressure
\begin{flalign}
\label{supS}
&\sigma=-\frac{1}{4\pi x}\bigg(\sqrt{1-\kappa r_{\Sigma}^2}-\dot\tau\sqrt{\frac{f}{g}}\bigg),\\
\label{pre}
&p=\frac{1}{8\pi x^2}\bigg([K_{\theta\theta}]+x^2K_{tt}^-\bigg).
\end{flalign}

Substituting condition ($\ref{const}$) and the metric components ($\ref{metricS}$) into \cref{supS}, we obtain, for a spatially flat universe ($\kappa=0$), the modified version of the Friedmann equation 
\begin{equation}\label{supS1}
H^2 = \frac{x_{0}+2m}{x^3} - \frac{2mx_{0}}{x^4} + \frac{8\pi\sigma}{x} + 16\pi^2\sigma^2,
\end{equation}
where $H\equiv\dot a/a$ is the Hubble parameter. The continuity equation is given by the following relation (see Appendix \ref{apd:A} for details)
\begin{equation}\label{conser}
\dot\sigma+2H(\sigma+p)=Q,
\end{equation}
where the energy flux term is given by
\begin{equation}\label{flux}
Q\equiv-\frac{(fg)'}{fg}\frac{H}{8\pi}(1+4\pi x\sigma).
\end{equation}
In the limit $x\gg x_0$, $fg  \rightarrow 1$ and $Q  \rightarrow 0$.

Note also that, although the matching with the expanding FLRW is formulated using the black hole's static external metric, the cosmological dynamics remains unchanged when the black hole's interior metric, which is isometric to a homogeneous and anisotropic Kantowski-Sachs space-time, is employed in the region $x_0 < x < 2m$. In other words, the dynamical hypersurface $\Sigma$ can continuously traverse the region between the transition surface and the horizon, keeping the modified equations invariant. It is worth mentioning that, although we have obtained the modified Friedmann equation from the standard DI matching conditions, in Appendix \ref{apd:B} we show that \cref{supS1} is consistent with the Hamiltonian formulation in the presence of a thin shell \cite{HA, FR}.  

From the modified Friedmann equation, we directly obtain the corresponding acceleration equation by differentiating with respect to cosmic time, yielding
\begin{eqnarray}
\nonumber
\frac{\ddot x}{x} &=& \frac{2mx_{0}}{x^4} -\frac{x_{0}+2m}{2x^3}
+\frac{4\pi\sigma}{x}+\\
\label{sup1}
&&\quad\quad\quad\quad\quad+16\pi^2\sigma^2  +\frac{4\pi\dot\sigma}{xH}\left(1+4\pi x\sigma\right).
\end{eqnarray}
In the limit $x\rightarrow x_0$, \cref{metric1,metricS}, together with \cref{supS}, yield a negative surface energy density,
\begin{equation}\label{en}
\sigma_B=-\frac{1}{4\pi x_0}.
\end{equation}
Substituting $\sigma_B$ into the modified Friedmann equation ($\ref{supS1}$) we obtain $H(x_0)=0$.Evaluating the acceleration equation at $x_0$ results in
\begin{equation}
\frac{\ddot{a}}{a}\bigg|_{B} = \frac{2m-x_0}{2x_0^3}.
\end{equation}
Since the transition surface lies inside the black-hole horizon \cite{Fernando}, $2m>x_0$, the acceleration at $x=x_0$ is positive, $\ddot a|_B>0$. Therefore, the scale factor reaches a local minimum, indicating that the universe undergoes a quantum bounce at $x=x_0$.

\section{Effective fluid and scalar field mapping \label{secIV}}

When matching the metric ($\ref{metricS}$) with an external FLRW universe, the extrinsic curvature undergoes a jump across the hypersurface $\Sigma$. According to the DI conditions, this jump generates a thin shell characterized by a surface energy density $\sigma$ and a surface pressure $p$. We can then extend this construction by cutting out an infinite number of non-overlapping spherical regions from the FLRW spacetime, enclosed by its respective thin shell, and inserting these LQG-corrected black hole spacetimes. The enclosing hypersurfaces are comoving with the FLRW cosmological expansion and share the same total energy as the excised background regions \cite{K, SW,G,F}. Therefore, this construction can be regarded as a variant of the Swiss-cheese model.

In this framework, we consider a universe filled with a homogeneous and isotropic distribution of black holes. The expanding universe then responds to an effective fluid arising from the dynamical shells, the contribution of the black-hole mass $m$, and the associated quantum corrections.
In this way, we can write the modified Friedmann equation ($\ref{supS1}$) in the form
\begin{equation}\label{H}
H^2=\frac{8\pi}{3}\bigg[\rho_m\bigg(1-\frac{x_0}{x}\bigg)+\rho_q+\rho_{\sigma}\bigg],
\end{equation}
where we defined the energy densities $\rho_m\equiv 3m/4\pi x^3$ and $\rho_q \equiv 3x_0/8\pi x^3$, with constant parameter $m$, and
\begin{equation}\label{den}
\rho_{\sigma}\equiv 6\pi\sigma^2+\frac{3\sigma}{x}
\end{equation}
stands for the effective energy density associated with the surface energy.
Moreover, surface energy $\sigma$ and pressure $p$ induce an effective pressure
\begin{equation}\label{pre}
P_\sigma = 2\pi \sigma^2 + 8\pi \sigma p + \frac{2p}{x}, \end{equation}
which ensures the continuity equation
\begin{equation}\label{cont1}
\dot\rho_{\sigma}+3H(\rho_{\sigma}+P_{\sigma})=\frac{3}{x}(1+4\pi x\sigma)Q.
\end{equation}

\subsection{Shell scenarios (de Sitter and dust)}

In this subsection, we evaluate two cases for the shell's equation of state (EoS): one representing a vacuum-type EoS and the other, pressureless dust. Based on these reference cases, we generalize the analysis in the following subsection to a constant equation of state parameter, which allows us to construct a representation of the scalar field and its dynamics throughout the cosmic expansion.

We first consider a vacuum-like EoS $p=-\sigma$. In this case, the continuity equation ($\ref{conser}$) provides the solution 
\begin{equation}\label{densi}
\sigma(x)=-\frac{1}{4\pi x_0}\left(2-\frac{x_0}{x}\right),
\end{equation}
where the integration constant is set to zero to be consistent with the bounce condition ($\ref{en}$), and for $x\gg x_0$ we have $\sigma\rightarrow2\sigma_B.$ In the asymptotic limit ($x\to\infty$), the Hubble rate given by equation ($\ref{H}$) is dominated by the last term, so we find the solution
\begin{equation}
x(t)=x_0\rm{exp}({8\pi\lvert\sigma_B\rvert\ t}).
\end{equation}
On the other hand, from the effective fluid viewpoint, using $p = -\sigma$ in equation ($\ref{pre}$) together with equation ($\ref{den}$) results in an asymptotic de Sitter phase with
\begin{equation}
P_\sigma\approx-\rho_{\sigma}.
\end{equation}

Another scenario is presented by a shell filled with a pressureless fluid $(p=0)$, whose continuity \cref{conser} provides the following solution for the surface energy density
\begin{equation}
    \sigma(x)=-\frac{x_0}{4\pi x^2}+\frac{x_0 \sqrt{x-x_0}}{4\pi x^{5/2}}\ln\left[\sqrt{\frac{x}{x_{0}}-1}+\frac{x}{x_{0}}\right].
\end{equation}
By setting $p = 0$ in effective pressure $(\ref{pre})$, and taking the limit $6\pi\sigma^2 \gg|3\sigma/x|$ of \cref{den,pre}, the effective fluid asymptotically exhibits a radiation-like behavior, with EoS
\begin{equation}
P_\sigma \approx \frac{1}{3}\rho_\sigma. \label{S46}
\end{equation}

\begin{figure*}[!t]
\centering
\includegraphics[width=0.5\linewidth]{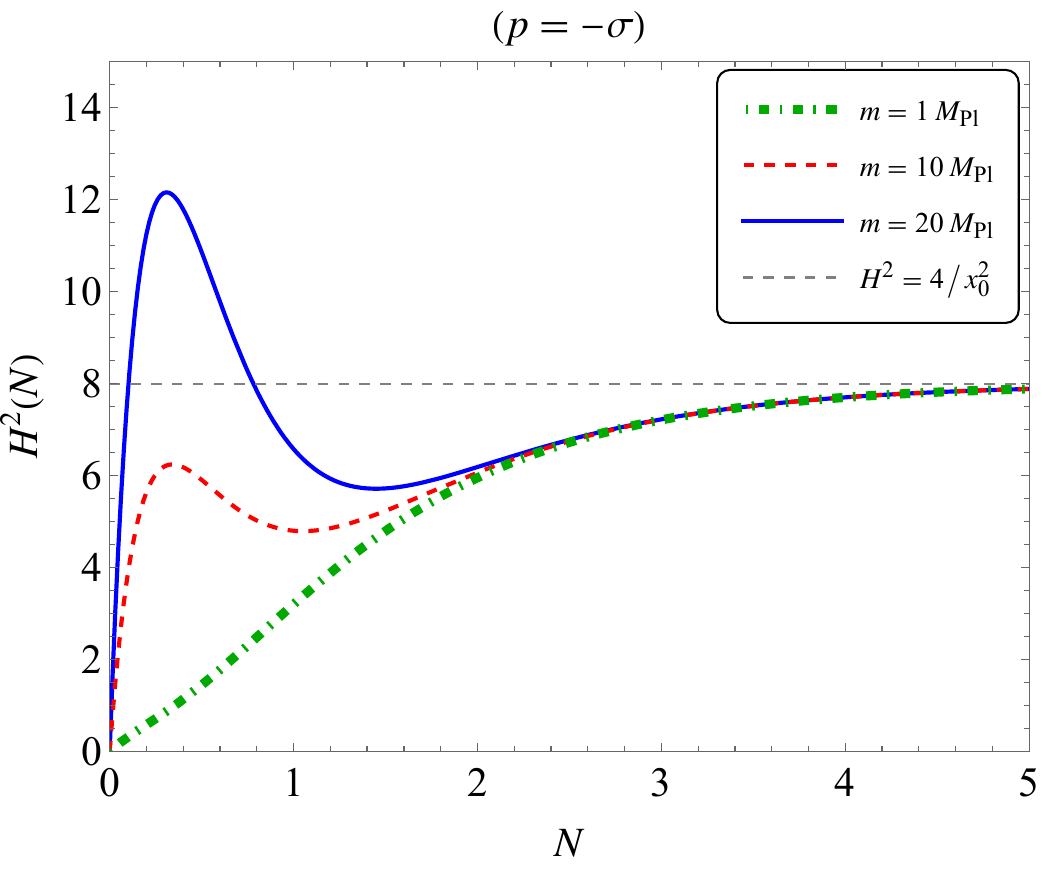}%
\includegraphics[width=0.5\linewidth]{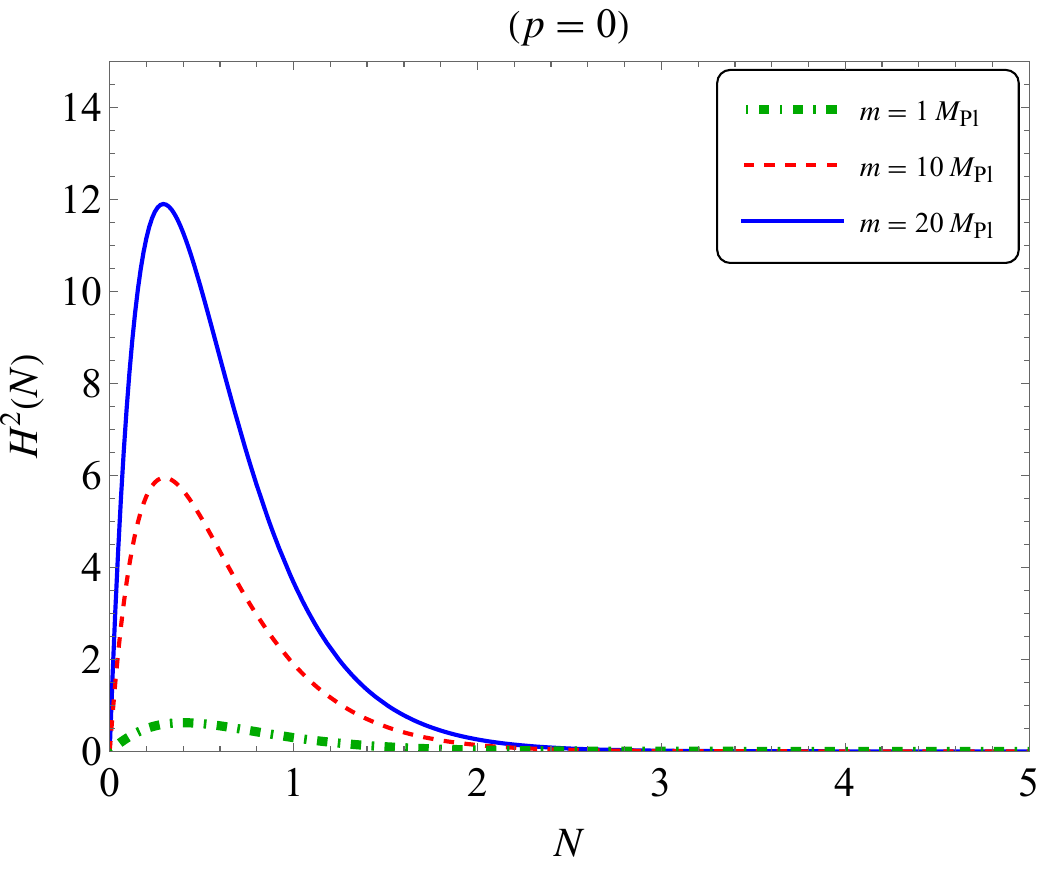}
\caption{\textit{Left panel:} The evolution of $H^2$ as a function of the number of $e$-folds $(N)$ for a vacuum-like shell for three different values of the black hole mass parameter $m$. \textit{Right panel:} We plot $H^2(N)$ for a dust-like shell.}
\label{fig:H2}
\end{figure*}

In \cref{fig:H2}, we show  the evolution of \eqref{H}  as a function of the number of $e$-folds $N \equiv \ln(x/x_0)$ for the two scenarios described above, for different values of the mass parameter $m$. At the left panel, for EoS    $p=-\sigma$, the bounce, where $H=0$, is followed by rapid expansion, until   $H^2$ reaches a peak, with its amplitude and duration being sensitive to the black hole mass parameter $m$, with larger masses producing higher peaks. As expansion proceeds, the universe settles onto the de Sitter plateau, where $H^2 \to 4/x_0^2$.  In the right panel, for the dust scenario ($p = 0$), after reaching the peak, $H^2$ goes to zero. In \cref{decay} we will show that this decay in $H^2$ is always true, provided that $\omega > -1$. 

\subsection{Single scalar field}

Since the effective fluid has well-defined energy density and pressure, this allows us to investigate whether it admits an equivalent scalar field description. Indeed, the energy density and pressure of a spatially homogeneous canonical scalar field can be expressed in terms of its kinetic energy and potential as
\begin{flalign}
\label{den1}
\rho_{\phi} = \frac{1}{2}\dot{\phi}^2 + V(\phi),\\
\label{pre1}
P_{\phi} = \frac{1}{2}\dot{\phi}^2 - V(\phi).
\end{flalign}
The appropriate correspondence is achieved by mapping the effective density and pressure, given by \cref{den,pre}, onto the scalar field energy density and pressure $(\ref{den1})$ and $(\ref{pre1})$. We can therefore determine the kinetic energy and the scalar potential, without making any assumption about $Q$, given by
\begin{flalign}
\label{pi}
&\dot{\phi}^2(x)=8\pi\sigma(p+\sigma)+\frac{1}{x}(3\sigma+2p),\\
\label{v1}
&V(x)=2\pi\sigma(\sigma-2p)+\frac{1}{2x}(3\sigma-2p).
\end{flalign}
From the above expressions and the continuity equation ($\ref{cont1}$), we verify that the effective field satisfies a modified Klein-Gordon equation with a source term
\begin{equation}\label{KG}
\ddot{\phi} + 3H\dot{\phi} + \frac{\partial V}{\partial\phi} = \frac{3(1 + 4\pi x\sigma)}{x \dot{\phi}}Q,
\end{equation}
according to \cref{cont1}. Therefore, the thermodynamic properties of the thin shells effectively act as a scalar field. 

\subsubsection{Transient phantom-like behavior near the bounce}
\label{phantom}
From this point forward, we  consider an EoS $p = \omega \sigma$, where $\omega$ is the shell equation of the state parameter. Using this EoS and surface energy density ($\ref{en}$) in Eqs. ($\ref{pi}$) and ($\ref{v1}$), we obtain the initial conditions
\begin{equation}
\dot{\phi}^2(x_0) = V(x_0) = -\frac{1}{4\pi x_0^2}.
\label{eq:initial_conditions}
\end{equation}
Furthermore, we can use $\phi = i\chi$ such that the real field acquires a sign reversal in its kinetic term near the bounce, exhibiting a local phantom-like behavior. It should be noted that this behavior is
transient for $\omega>-1$. In fact, from Eq. ($\ref{pi}$) the kinetic term vanishes at a radius $x_t$, and for $x > x_t$ the field  recovers its canonical nature.

Additionally, in the vicinity of $x_0$ the Hubble equation ($\ref{H}$) can be approximated by the expression $H^2\approx (2m - x_0)(x-x_0)x_0^{-4}$, up to first order. Then, substituting this result and the initial conditions ($\ref{eq:initial_conditions}$) into the kinetic term given by \cref{pi}, we obtain the solution for the real field
\begin{equation}
\label{chi}
\chi(x) \approx \chi_0 \pm \sqrt{\frac{x - x_0}{\pi (2m - x_0)}}. 
\end{equation}
On the other hand, it can be shown that the surface energy density, governed by the continuity equation ($\ref{conser}$), can be approximated by the expression
\begin{equation}\label{surfac}
\sigma(x) \approx -\frac{1}{4\pi x_0} + \left[\frac{4\omega + 3}{4\pi x_0^2}\right](x - x_0).
\end{equation}
Expanding the effective potential \eqref{v1} near $x_0$ and then using \cref{surfac,chi}, we find 
\begin{equation}
\label{S55}
V(\chi) \approx -\frac{1}{4\pi x_0^2} + \frac{M^2}{2}(\chi - \chi_0)^2, 
\end{equation}
where $M = \sqrt{(4\omega^2+4\omega+3)(2m-x_0)/2x_0^3}$ is the mass of the scalar field. Since $2m > x_0$, we obtain a local minimum for the potential at the bounce, so that the field climbs up the potential as the universe expands.

\subsubsection{Asymptotic decay}\label{decay}

To analyze the expansion of the universe in the regime far from the bounce ($x \gg x_0$), where the energy flux $Q$ vanishes, we consider the Klein-Gordon equation ($\ref{KG}$). In this regime, the continuity equation ($\ref{conser}$) reduces to
\begin{equation}
\dot{\sigma} + 2(1 + \omega)H\sigma = 0,
\label{continuity}
\end{equation}
whose solution is given by
\begin{equation}
\sigma(x) = -\frac{1}{2\pi x_0} \left( \frac{x}{x_*} \right)^{-2(1+\omega)},
\label{sigma}
\end{equation}
where $x_*$ is an integration constant. 

Moreover, \cref{pi,v1} can be approximated by
\begin{eqnarray}
\label{eq:phidot_approx_asymp}
\dot{\phi}^2 &\approx& 8\pi \sigma^2 (1 + \omega),\\ 
\label{V_sigma}
V&\approx& 2\pi \sigma^2 (1 - 2\omega).
\end{eqnarray}
Since the Friedmann equation during this stage is dominated by the effective energy density $\rho_\sigma \approx 6\pi \sigma^2$, the Hubble rate satisfies $H^2 \approx 16\pi^2 \sigma^2$. Combining this with \cref{eq:phidot_approx_asymp} yields
\begin{equation}
\frac{d\phi}{dx}\approx \sqrt{\frac{1 + \omega}{2\pi}} \frac{1}{x}.
\label{eq:dphidx_reheating}
\end{equation}
Integrating the above equation results in the logarithmic evolution of the scalar field
\begin{equation}\label{ra}
\phi(x) \approx \phi_* + \sqrt{\frac{1 + \omega}{2\pi}} \ln \left( \frac{x}{x_{\text{*}}} \right).
\end{equation}
Substituting the surface energy density (\ref{sigma}) and \cref{ra} into the potential (\ref{V_sigma}), we obtain the evolution of the effective potential as
\begin{equation}
V(\phi) \approx \frac{(1-2\omega)}{2\pi x_0^2} e^ {-4 \sqrt{2\pi (1 + \omega)} (\phi - \phi_{\mathrm{*}})}. \label{S65}
\end{equation}

Along these lines, the effective energy density $\rho_\phi$ and pressure $P_\phi$ given, respectively, by \cref{den,pre} reduce to
\begin{flalign}
&\rho_{\phi}\approx 6\pi\sigma^2,\\
\label{p1}
&P_\phi \approx 2\pi \sigma^2 (1 + 4\omega), 
\end{flalign}
whose ratio $P_\phi / \rho_\phi$ provides
\begin{equation}
\omega_\phi \approx\frac{1 + 4\omega}{3}.
\label{eq:w_phi_relation}
\end{equation}
It is straightforward to verify that for $\omega>-1$, the scalar potential decays exponentially as the universe expands. Conversely, for $\omega = 0$, the effective scalar field behaves asymptotically as radiation.

\subsection{Two scalar field mechanism for inflation}
\label{sec:two_field}

As shown in \cref{phantom}, a characteristic feature of the field near the bounce is the temporary  negative sign of its kinetic energy term. The bounce occurs at the Planck scale, setting the initial conditions for the evolution of the early universe with the field emerging from the bottom of a parabolic potential. However, to connect with the observed universe, this Planckian energy density must be suppressed. Notably, for any $\omega > -1$, it was shown that the effective potential decays exponentially in the regime $x\gg x_0$, leading the energy density to dilute as $\rho_\phi \propto a^{-3(1+\omega_\phi)}$. To reconcile this high-energy bounce with a subsequent inflationary phase, we introduce a second scalar field whose potential exhibits a plateau at an energy scale constrained by CMB observations. To realize this two-stage scenario, we decompose the surface energy density $\sigma$ and the pressure $p$ of the shell as
\begin{flalign}
\label{eq:sigma_decomp}
&\sigma = \sigma_{1} + \sigma_{2},\\
\label{eq:p_decomp}
&p = p_{1} + p_{2}, 
\end{flalign}
where the first component satisfies the EoS $p_1 = \omega \sigma_1$, and the condition $|\sigma_{1}| \gg \sigma_{2}$ holds near $x_0$.  With this decomposition, the potential (\ref{v1}) can be recast as two interacting homogeneous scalar fields with potential
\begin{equation}
V(\phi,\varphi)=V(\phi)+V(\varphi)+V_{int},
\end{equation}
where the interaction potential is given by
\begin{equation}\label{int}
V_{int} = 4\pi(\sigma_{1} \sigma_{2}-\sigma_{1} p_{2} - \sigma_{2} p_{1}).
\end{equation}
The same procedure can be applied to the kinetic term ($\ref{pi}$). Assuming the EoS $p_2 \approx -\sigma_2$ for the second field during its quasi-de Sitter phase, the interaction potential in \cref{int} reduces to
\begin{equation} 
V_{\text{int}} \approx 4\pi \sigma_1 \sigma_2 (2 - \omega).
\end{equation}

In this two-stage scenario, the scalar field $\phi(t)$ governs the primordial dynamics in the vicinity of $x_0$. As the universe expands, its energy density $\rho_{\phi}$ dilutes according to $\omega_{\phi} > -1$. Once $\rho_{\phi}$ falls below the inflation energy scale, the potential $V(\varphi)$ of the second scalar field, $\varphi(t)$, gradually dominates and drives the slow-roll inflation.

\begin{figure}[htbp]
    \centering    \includegraphics[width=0.5\textwidth]{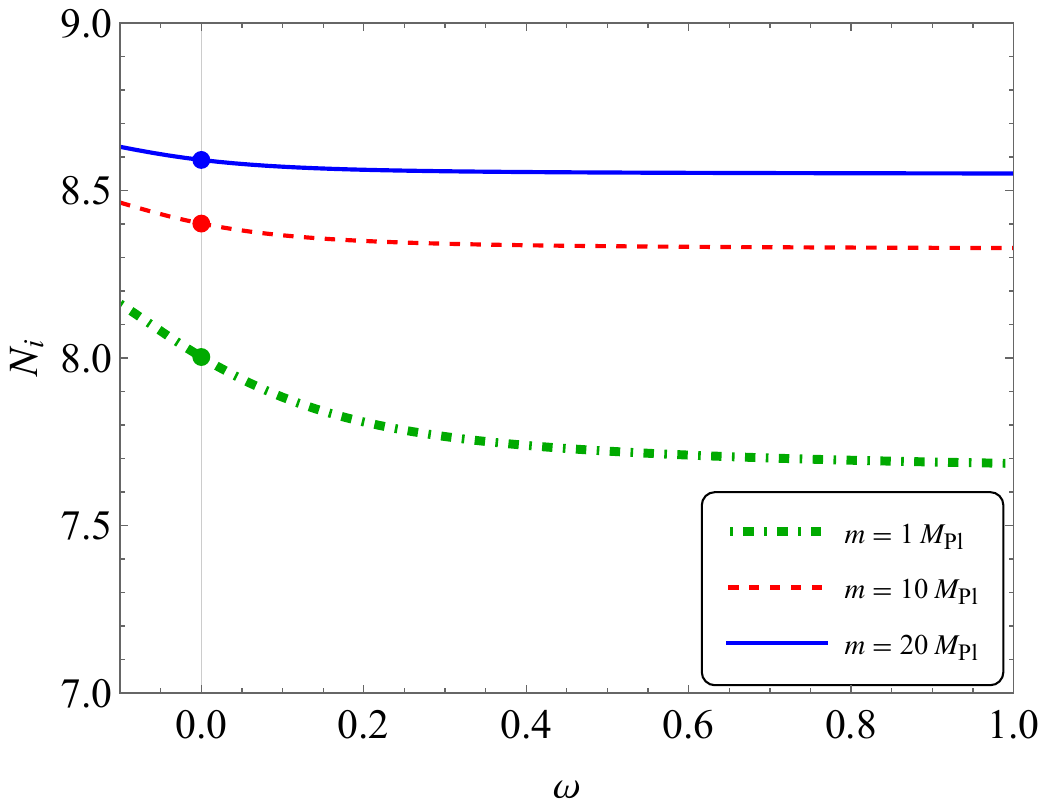} 
    \caption{Number of $e$-folds $N_\text{i}$ as a function of $\omega$ for three different values of the mass parameters $m$.}
    \label{fig:Nw}
\end{figure}

Therefore, to quantify the duration of this pre-inflationary stage, we evaluate the total number of $e$-folds $N_{\text{i}} \equiv \ln(x_{\text{i}}/x_0)$ required for the Hubble rate to drop from the Planckian regime to the inflation energy scale $\rho_{\text{i}} \sim 10^{-10} M_{\text{Pl}}^4$. In \cref{fig:Nw}, we show the numerical evolution of $N_{\text{i}}$ as a function of the EoS parameter $\omega$ in the range $[-0.1, 1.0]$ for three values of the mass parameters $m$. Throughout this range, the duration of the pre-inflationary stage requires only $N_{\text{i}} \approx 7.7 - 8.6$ $e$-folds to reach the inflation threshold, confirming that this initial phase is brief and paves the way for the $\sim 50-60$ $e$-folds of standard inflation.
 
\section{CONCLUDING REMARKS \label{secV}}

In this work, we investigate a non-singular primordial cosmology resulting from matching an LQG-inspired effective black hole geometry with an FLRW universe metric via DI junction conditions. The resulting junction conditions lead to modified Friedmann and shell continuity equations that replace the initial singularity with a quantum bounce at a minimum radius $x_0$.

By mapping the thin-shell surface energy-momentum tensor to a homogeneous scalar field, we reconstruct its primordial dynamics. In the vicinity of the bounce, the kinetic term is negative, giving the scalar field a transient phantom-like behavior. Thus, the field climbs a parabolic potential, initiating a high-energy accelerated regime.
The asymptotic behavior of the model reveals a dependence on the shell's equation of state. While a parameter $\omega = -1$ drives the universe to a de Sitter phase, we show that for any equation of state satisfying $\omega > -1$, the effective energy density of the field is rapidly suppressed as it becomes canonical.

To bridge the energy gap between the Planckian bounce and the sub-Planckian scale of slow-roll inflation, we proposed a two-component decomposition of surface energy density and surface pressure. By solving the resulting continuity equation, including energy flux corrections, we show that the duration of the pre-inflationary stage is brief.  In the parameter space $\omega \in [-0.1, 1.0]$ and $m \in [1, 20] M_\text{Pl}$, the duration of this pre-inflationary stage remains brief, requiring only $N_{\text{i}} \approx 7.7 - 8.6$ e-folds for the Hubble rate to drop from the Planck scale to the inflation threshold. While the minimum radius $x_0 = 1/\sqrt{2} \approx 0.71 \, \ell_{\text{Pl}}$ is fixed by the LQG area gap (for $\gamma = \sqrt{3}/6$), ensuring a Planckian bounce without fine-tuning, the black hole mass scale $m \sim \mathcal{O}(1-20)\,M_{\text{Pl}}$ represents a choice of the primordial mass. This demonstrates the robustness of the dilution of energy density, paving the way for the second scalar field to take over and drive the $50$--$60$ $e$-folds of standard slow-roll inflation required by CMB observations.

The next step is to calculate the primordial cosmological perturbations generated during the bounce and subsequent evolution. This analysis will determine the primordial scalar and tensor power spectra, providing predictions for possible observational signatures of the pre-inflationary bounce at large angular scales of the CMB.

\begin{acknowledgments}
We thank Saulo Carneiro for a first reading. VFCB, FGSM and SMR thank FAPESB (Brazil) for financial support under the grant numbers BOL1283/2025, BOL0773/2025 and BOL0867/2024, respectively.
\end{acknowledgments}

\appendix
\section{Derivation of the continuity equation \label{apd:A}}

\label{app:continuity}

The surface energy density ($\ref{supS}$) can be written as
\begin{equation}\label{suf}
\sigma = -\frac{1}{4\pi x} (1 - \alpha_-),
\end{equation}
where the factor $\alpha_-$ is described by the LQG black hole metric \eqref{metricS}
\begin{equation}
\alpha_- \equiv \sqrt{\frac{1}{g(x)} + \dot{x}^2}.
\end{equation}
Differentiating $\sigma$ with respect to cosmological time, we obtain
\begin{equation}
\dot{\sigma} + 2H\sigma = H\sigma + \frac{\dot{\alpha}_-}{4\pi x}.
\end{equation}

Combining the temporal component of the extrinsic curvature $(\ref{kt})$  and surface pressure ($\ref{pre}$), we find
\begin{equation}
\frac{\dot{\alpha}_-}{4\pi x} = -2Hp - H\sigma - \frac{(fg)'}{fg} \frac{H \alpha_-}{8\pi}.
\end{equation}
Substituting this back into ($\ref{suf}$) yields
\begin{equation}
\dot{\sigma} + 2H(\sigma + p) = -\frac{(fg)'}{fg} \frac{H}{8\pi} (1 + 4\pi x \sigma).
\end{equation}

\section{Matching conditions via Hamiltonian formalism \label{apd:B}}

In addition to the standard DI junction formalism, the same dynamics presented in this work can be obtained through a Hamiltonian formulation. By introducing a foliation of spacetime into ADM spatial slices $\Sigma_t$ that intersect the timelike hypersurface $\Sigma$, the Hamiltonian constraint can be integrated across a radial infinitesimal length
centered in the shell position $x=R$ at a constant time to obtain its equation of motion. The surface energy-momentum tensor of the fluid is localized on the hypersurface $\Sigma$ through a Dirac delta distribution
\begin{equation}
T_{\mu\nu} = S_{\mu\nu} \delta(X),
\end{equation}
where $S_{\mu\nu}$ is the four-dimensional version of the surface stress-energy tensor given by equation ~(\ref{perf}), and $X$ is the proper distance along the spacelike normal vector $\xi^\mu$. The matter contribution in the Hamiltonian constraint is given by $T_{\perp\perp} = T_{\mu\nu} n^\mu n^\nu$, where $n^\mu$ is the unit normal vector to the spatial slices $\Sigma_t$. We can decompose $n^\mu$ as
\begin{equation}
n^\mu = -u_{\perp} u^\mu + \xi_{\perp} \xi^\mu,
\end{equation}
where $u_\perp \equiv u^\mu n_\mu$, $\xi_\perp \equiv \xi^\mu n_\mu$ and $u^\mu \xi_\mu = 0$. The normalization $n^{\mu}n_{\mu}=-1$ yields the identity
\begin{equation}
u_{\perp}^2 - \xi_{\perp}^2 = 1.
\label{eq:norm_identity}
\end{equation}
Then, the stress-energy tensor is found to be
\begin{equation}
T_{\perp\perp} = \sigma u_{\perp}^2 \delta(X),
\end{equation}
where we have used $h_{\mu\nu} n^\mu n^\nu = (g_{\mu\nu} - \xi_{\mu}\xi_{\nu})n^\mu n^\nu = -1 - \xi_{\perp}^2$ and \cref{eq:norm_identity}, so that the pressure does not contribute to the Hamiltonian constraint.

In terms of a general metric component written in standard spherical coordinates,
 we find $u_{\perp}^2 = g(x) \alpha^2$, where $\alpha \equiv \sqrt{(g_{xx})^{-1} + \dot{x}^2}$. In a constant-time slice $\Sigma_t$, the transformation of the Dirac delta distribution from the proper normal coordinate $X$ to the spatial coordinate $x$ carries the Jacobian $dX/dx = g_{xx}\alpha$, so that $\delta(X) = \frac{1}{g_{xx}\alpha} \delta(x - R)$. Then, the projected stress-energy tensor simplifies to
\begin{equation}
T_{\perp\perp} = \alpha \sigma \delta(x - R).
\end{equation}

In the ADM formulation, the total Hamiltonian constraint satisfies
\begin{equation}
\mathcal{H}_\perp = \mathcal{H}_\perp^{\text{grav}} + \sqrt{h} T_{\perp\perp} \approx 0.
\end{equation}
The classical gravitational Hamiltonian constraint takes the explicit form \cite{HA}
\begin{equation}
\mathcal{H}_\perp^{\text{grav}} = -\frac{\sqrt{h}}{8\pi } \left[ \frac{1}{x^2}\left(1 - \frac{1}{g(x)}\right) - \frac{1}{x}\left(\frac{1}{g(x)}\right)' \right],
\end{equation}
where $h \equiv \det(g_{ij})$ denotes the determinant of the spatial metric and the prime denotes the derivative with respect to the radial coordinate $x$. For the Schwarzschild metric we obtain $\mathcal{H}_\perp^{\text{grav}} \approx 0$, reflecting the fact that the Schwarzschild metric is a vaccum solution in GR. However, when evaluating $\mathcal{H}_\perp^{\text{grav}}$ using the metric function $g^{-1}(x) = (1 - 2m/x)(1 - x_0/x)$, we find a non-zero contribution
\begin{equation}
\mathcal{H}_\perp^{\text{grav}} = -\frac{\sqrt{h}}{8\pi} \left( \frac{2m x_0}{x^4} \right).
\end{equation}
Consequently, when evaluated within the classical canonical constraint, the quantum spacetime geometry manifests as an effective quantum energy density.

Now, we integrate the total constraint $\mathcal{H}_\perp \approx 0$ across an infinitesimal radial interval $[x - \varepsilon, x + \varepsilon]$ centered at the shell position $x$ on the constant-time slice $\Sigma_t$ \cite{HA, FR}. In the limit $\varepsilon \to 0$ we obtain the jump condition
\begin{equation}
\alpha_+ - \alpha_- = -4\pi  \sigma x,
\label{eq:jump_condition}
\end{equation}
where, on the exterior side described by the FLRW metric, the factor $\alpha_+$ is given by
\begin{equation}
\alpha_+ \equiv \sqrt{1 - \kappa r_\Sigma^2}.
\end{equation}
For a spatially flat universe ($\kappa = 0$), substituting $\alpha_+$ and $\alpha_-$ into \cref{eq:jump_condition}, we recover the modified Friedmann equation \eqref{supS1}.

\bibliographystyle{apsrev4-2}
\bibliography{refs}

\end{document}